\documentclass[aps,pre,a4paper,10pt,twocolumn,groupedaddress]{revtex4-1} 
\usepackage{amsmath} 
\usepackage{amssymb} 
\usepackage{bbm} 
\usepackage{graphicx} 
\usepackage{epstopdf} 
\usepackage{color}
\usepackage{hyperref}

\begin{document} 
%%%%%%%%%%%%%%%%%%%%%%%%%%%%%%%%%%%%%%%%%%%%%%%%%%%%%%%%%%%%%%%%%%%%%%%%%%%%%%%%%%%%%%%%%%%%%%%%%%%%%%%%%%% 
%%%%%%%%%%%%%%%%%%%%%%%%%%%%%%%%%%%%%%%%%%%%%%%%%%%%%%%%%%%%%%%%%%%%%%%%%%%%%%%%%%%%%%%%%%%%%%%%%%%%%%%%%%% 
%%%%%%%%%%%%%%%%%%%%%%%%%%%%%%%%%%%%%%%%%%%%%%%%%%%%%%%%%%%%%%%%%%%%%%%%%%%%%%%%%%%%%%%%%%%%%%%%%%%%%%%%%%% 
%%%%%%%%%%%%%%%%%%%%%%%%%%%%%%%%%%%%%%%%%%%%%%%%%%%%%%%%%%%%%%%%%%%%%%%%%%%%%%%%%%%%%%%%%%%%%%%%%%%%%%%%%%% 
\title{Thermally induced passage and current of particles in a highly unstable optical potential} 
%%%%%%%%%%%%%%%%%%%%%%%%%%%%%%%%%%%%%%%%%%%%%%%%%%%%%%%%%%%%%%%%%%%%%%%%%%%%%%%%%%%%%%%%%%%%%%%%%%%%%%%%%%% 
%%%%%%%%%%%%%%%%%%%%%%%%%%%%%%%%%%%%%%%%%%%%%%%%%%%%%%%%%%%%%%%%%%%%%%%%%%%%%%%%%%%%%%%%%%%%%%%%%%%%%%%%%%% 
\author{Artem Ryabov} 
\email{rjabov.a@gmail.com} 
\affiliation{Charles University in Prague, Faculty of Mathematics and Physics, Department of Macromolecular Physics, V Hole{\v s}ovi{\v c}k{\' a}ch~2, 180~00~Praha~8, Czech Republic} 
\author{Pavel Zem{\' a}nek} 
\email{zemanek@isibrno.cz} 
\affiliation{Institute of Scientific Instruments of CAS, Czech Academy of Sciences,
Kr{\' a}lovopolsk{\' a}~147, 612~64~Brno, Czech Republic} 
\author{Radim Filip} 
\email{filip@optics.upol.cz} 
\affiliation{Department of Optics, Palack{\' y} University, 17.~listopadu~1192/12, 771~46~Olomouc, Czech Republic}

%%%%%%%%%%%%%%%%%%%%%%%%%%%%%%%%%%%%%%%%%%%%%%%%%%%%%%%%%%%%%%%%%%%%%%%%%%%%%%%%%%%%%%%%%%%%%%%%%%%%%%%%%%%
%%%%%%%%%%%%%%%%%%%%%%%%%%%%%%%%%%%%%%%%%%%%%%%%%%%%%%%%%%%%%%%%%%%%%%%%%%%%%%%%%%%%%%%%%%%%%%%%%%%%%%%%%%%
\date{\today} 
%%%%%%%%%%%%%%%%%%%%%%%%%%%%%%%%%%%%%%%%%%%%%%%%%%%%%%%%%%%%%%%%%%%%%%%%%%%%%%%%%%%%%%%%%%%%%%%%%%%%%%%%%%%
%%%%%%%%%%%%%%%%%%%%%%%%%%%%%%%%%%%%%%%%%%%%%%%%%%%%%%%%%%%%%%%%%%%%%%%%%%%%%%%%%%%%%%%%%%%%%%%%%%%%%%%%%%%
\begin{abstract} 
We discuss the statistics of first-passage times of a Brownian particle moving in a highly unstable nonlinear potential proportional to an odd power of position. We observe temperature-induced shortening of the mean first-passage time and its dependence on the power of nonlinearity. We propose a passage-time fraction as both a simple and experimentally detectable witness of the nonlinearity. It is advantageously independent of all other parameters of the experiment and observable for a small number of trajectories. To better characterize the stochastic passage in the unstable potential, we introduce an analogy of the signal-to-noise ratio for the statistical distribution of the first-passage times. Interestingly, the upper bound for the signal-to-noise ratio is temperature independent in the unstable potential. Finally, we describe the nonequilibrium steady state of the particle cyclically passing through unstable odd nonlinearity. The maximum of the steady-state probability distribution shifts against the directions of the current and this counterintuitive effect increases with temperature. All these thermally induced effects are very promising targets for experimental tests of highly nonlinear stochastic dynamics of particles placed into optical potential landscapes of shaped optical tweezers.    
\end{abstract} 
%%%%%%%%%%%%%%%%%%%%%%%%%%%%%%%%%%%%%%%%%%%%%%%%%%%%%%%%%%%%%%%%%%%%%%%%%%%%%%%%%%%%%%%%%%%%%%%%%%%%%%%%%%%
%%%%%%%%%%%%%%%%%%%%%%%%%%%%%%%%%%%%%%%%%%%%%%%%%%%%%%%%%%%%%%%%%%%%%%%%%%%%%%%%%%%%%%%%%%%%%%%%%%%%%%%%%%%
% insert suggested PACS numbers in braces on next line
% \pacs{}
% insert suggested keywords - APS authors don't need to do this
%\keywords{}
%\maketitle must follow title, authors, abstract, \pacs, and \keywords
\maketitle
%%%%%%%%%%%%%%%%%%%%%%%%%%%%%%%%%%%%%%%%%%%%%%%%%%%%%%%%%%%%%%%%%%%%%%%%%%%%%%%%%%%%%%%%%%%%%%%%%%%%%%%%%%%
%%%%%%%%%%%%%%%%%%%%%%%%%%%%%%%%%%%%%%%%%%%%%%%%%%%%%%%%%%%%%%%%%%%%%%%%%%%%%%%%%%%%%%%%%%%%%%%%%%%%%%%%%%%

\section{Introduction}
%%%%%%%%% 
Recent progress in high-fidelity control and both fast and accurate measurements of a particle trapped in optical tweezers \cite{Berut2012, Blickle2012, Roldan2014, Quinto-Su2014, Li1673, Li2011, Gieseler2013, Gieseler2014, Kheifets1493, Romodina2015, Martinez2016} allows deeper experimental investigation of a broader class of transient effects in nonlinear stochastic dynamics. This experimental progress therefore stimulates new wave of theoretical investigations of highly nonlinear stochastic dynamics, especially close to strongly unstable points in the potential. The strong instability and nonlinearity can transfer an environmental thermal noise to useful positive effects like thermally-induced motion, oscillations and synchronization, the basic effects, which can serve as key ingredients of many and more complex phenomena occurring in nonlinear stochastic systems. Yet, their theoretical and experimental investigation is still far from being finished since the strong nonlinearity and significantly complicate statistical analysis of the dynamics. 
The most basic example is a transient short-time nonlinear behavior in the cubic potential which can transfer thermal noise from the environment to a displacement of the particle \cite{RadimPavelI}. This result and experimental possibilities to test it in optical tweezers \cite{RadimPavelII} motivate to examine the overdamped dynamics of particle passages and cyclic currents through instability of the cubic and higher-order potentials. In this case, more complex long-time dynamics can be potentially studied instead of the short-time dynamics, which opens a door to more experimentally accessible investigation of the thermally-induced nonlinear behavior.

In the present paper, we analyze the first-passage time statistics \cite{bookRedner, bookKrapivsky} of an overdamped particle diffusing in elementary unstable potentials $U(x)=\mu x^n$, $n=3,5,\ldots$ \cite{horsthemke89, Lindner03, Lindner04, ReimannPRL01, ReimannPRE02, Sancho91, ReimannBroeck94, Hirsch82, Pomeau80, Arecchi82, Vilar19991}. Instead of investigation of the statistical characteristics of the particle position, limited to short times by divergent trajectories \cite{RadimPavelI, RadimPavelII}, we concentrate on the discussion of complementary statistics of the particle passage time as the function of the position. In particular, we analyze the mean first-passage time focusing on inherently nonlinear effects induced by the temperature of the environment and introduce signal-to-noise ratio for the first-passage time statistics. We compare thermally-induced effects for elementary nonlinear potentials with the linear potential, corresponding to a constant force. This approach not only offers a complementary (global) information about the long-term dynamics inside the unstable potential, but it is also much simpler and reliable for the experimental verification of the nonlinear dynamics because it protects it from a strong divergence. We predict temperature-dependence of  the overall mean first-passage time between two symmetrical points (initial and final position) far enough from the center of the odd potential.  Simultaneously, we verify that the signal-to-noise ratio for the passage time becomes temperature-independent for a large distance of the points from the center. We introduce fractions formed by the mean first-passage times needed to pass through individual parts of the odd potential (from the center to minus infinity and from the plus infinity to the center). The fractions turn out to be very simple and directly measurable witnesses of the nonlinear stochastic dynamics since they depend on the order of the nonlinearity $n$ only. In the last part of the present paper, we extend the analysis to a current caused by a periodic passage assisted by a return mechanism moving the particle back from the final to the initial position. We observe a temperature-induced shift of a maximum of the steady-state position distribution against the probability current, i.e., the maximum shifts towards higher values of the potential. This counter-intuitive effect is simply proportional to $T^{1/n}$, where $T$ stands for the temperature. Our results can be directly experimentally verified using current laboratory technology with optical tweezers. After the experimental tests, this theoretical methodology can be applied to investigate much more complex nonlinear stochastic effects.

%%%%%%%%%%%%%%%%%%%%%%%%%%%%%%%%%%%%%%%%%%%%%%%%%%%%%%%%%%%%%%%%%%%%%%%%%%%%%%%%%%%%%%%%%%%%%%%%%%%%%%%%%%%
%%%%%%%%%%%%%%%%%%%%%%%%%%%%%%%%%%%%%%%%%%%%%%%%%%%%%%%%%%%%%%%%%%%%%%%%%%%%%%%%%%%%%%%%%%%%%%%%%%%%%%%%%%%
\section{HIGHLY UNSTABLE POTENTIALS AND EXPERIMENTAL SETTING}
%%%%%%%%%%%%%%%%%%%%%%%%%%%%%%%%%%%%%%%%%%%%%%%%%%%%%%%%%%%%%%%%%%%%%%%%%%%%%%%%%%%%%%%%%%%%%%%%%%%%%%%%%%%
%%%%%%%%%%%%%%%%%%%%%%%%%%%%%%%%%%%%%%%%%%%%%%%%%%%%%%%%%%%%%%%%%%%%%%%%%%%%%%%%%%%%%%%%%%%%%%%%%%%%%%%%%%%

Let us consider an overdamped Brownian motion in the potential $U(x)$,
\begin{equation}
\label{potentialU}
U(x) = \mu x^{n},\quad n\,\,  {\rm odd}. 
\end{equation} 
Thus the dynamics of the particle position $x(t)$ is described by the Langevin equation
\begin{equation}
\label{Langevin} 
\frac{\rm d }{{\rm d}t} x(t) =  \frac{1 }{\gamma} F\left( x(t) \right)  + \sqrt{2D} \xi (t),
\end{equation}
with $\xi(t)$ being the delta-correlated white noise, $\left<\xi(t)\right>=0$, $\left<\xi(t) \xi(t') \right> = \delta(t-t')$. The force $F(x)$ is determined by the externally imposed potential $U(x)$,  
$ 
F(x) = - {\rm d}U/{{\rm d} x}, 
$
and the diffusion constant, which controls the strength of the thermal noise, is given by the Einstein (fluctuation-dissipation) relation 
$D = {k_B T}/{ \gamma},$  
where $\gamma$ is the friction coefficient and $T$ stands for the temperature. 

For the odd orders $n$, $n\geq 3$, of this unstable potential, particle position reaches minus infinity, $x(t) \to - \infty$ {\em at a finite time} $\tau_\infty$ (we always assume that $\mu >0$).  When $D=0$, we have 
\begin{equation}
\label{tauinfty}	
\tau_{\infty}=-\frac{1}{\mu n(n-2)x(0)^{n-2}},\quad {{\rm for}\,\, x(0)<0}. 
\end{equation}	
%for $x(0)<0$. 
This fact restricts moment analysis of the stochastic dynamics limited to a short time only since the divergence of trajectories strongly affects the all statistical moments of position, see for example \cite{RadimPavelI}. However, even during this short-time dynamics both interesting and promising thermal-induced effects can be observed due to strong nonlinearity and instability.

We describe the random position $x(t)$ of the particle at a given time $t$ by the probability density function (PDF) $p(x,t|y)$, which is the conditional PDF for $x(t)$ given that, at $t=0$, the particle is located at $y$. Hence we have $p(x,0|y) = \delta(x-y)$. The latter equality serves as the initial condition for the \emph{backward} Fokker-Planck equation~\cite{bookGardiner}  
\begin{equation}
\label{BackwardFP} 
\frac{\partial }{\partial t} p(x,t|y) = 
D {\rm e}^{\beta U(y)} 
 \frac{\partial }{\partial y}
 \left[  
{\rm e}^{- \beta U(y)} 
\frac{\partial}{\partial y} p(x,t|y)\right],
\end{equation} 
where on the right-hand side occurs the adjoint Fokker-Planck operator. 
The backward equation (\ref{BackwardFP}) is a convenient tool for study of statistical properties of the \emph{first-passage} times. 

%%%%%%%%%%%%%%%%%%%%%%%%%%%%%%%%%%%%%%%%%%%%%%%%%%%%%%%%%%%%%%%%%%%%%%%%%%%%%%%%%%%%%%%%%%%%%%%%%%%%%%%%%%% 
%%%%%%%%%%%%%%%%%%%%%%%%%%%%%%%%%%%%%%%%%%%%%%%%%%%%%%%%%%%%%%%%%%%%%%%%%%%%%%%%%%%%%%%%%%%%%%%%%%%%%%%%%%% 
\begin{figure}[t!] 
\includegraphics[scale=0.75]{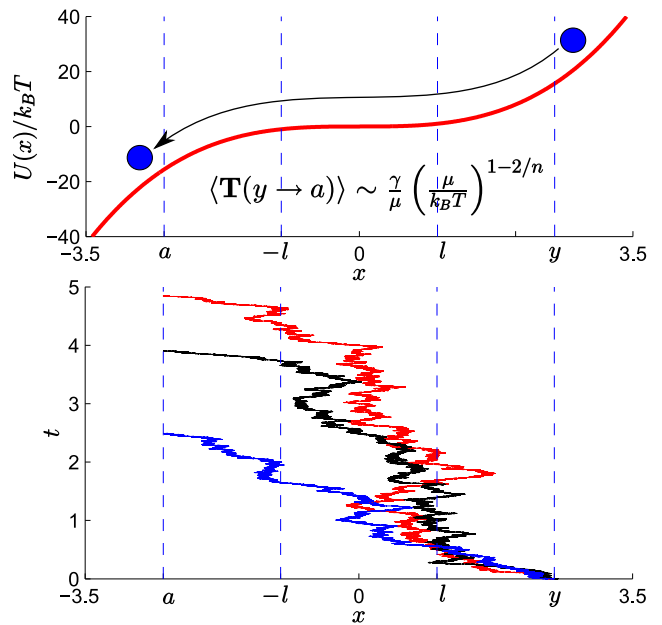}
\caption{\label{fig:schematic} 
Three trajectories of Brownian motion (the lower panel) in the nonlinear cubic potential $U(x)=\mu x^{n}$, $n=3$, schematically illustrated in the upper panel. Within the interval $(-l,l)$, $l=\left(k_B T/\mu  \right)^{1/n}$, cf.~Eqs.~(\ref{ldef}), thermal fluctuations have pronounced effect on the motion of the particle. The dynamics out of  $(-l,l)$ is rather fast and for $|x| \gg l$  it is close to the deterministic one since the thermal noise is negligible as compared to rapidly increasing (absolute) value of $U(x)$. The upper panel indicates also the scaling of mean first-passage time with the temperature $T$ as discussed in Sec.~\ref{subsec:nonlinear}.} 
\end{figure}  
%%%%%%%%%%%%%%%%%%%%%%%%%%%%%%%%%%%%%%%%%%%%%%%%%%%%%%%%%%%%%%%%%%%%%%%%%%%%%%%%%%%%%%%%%%%%%%%%%%%%%%%%%%%
%%%%%%%%%%%%%%%%%%%%%%%%%%%%%%%%%%%%%%%%%%%%%%%%%%%%%%%%%%%%%%%%%%%%%%%%%%%%%%%%%%%%%%%%%%%%%%%%%%%%%%%%%%% 

{\em Experimental setting} is sketched in Fig.~\ref{fig:schematic}. Assume that initially ($t=0$) the Brownian particle is located at the position $y$. The \emph{first-passage time} (FPT) $\mathbf{T}(y\to a)$ is simply the random time when the Brownian particle reaches the position $a$, $a\neq y$, for the first time. 
The experiment, where statistical properties of the first-passage time are measured, runs as follows. First, at $t=0$, we prepare the particle at a specified position $y$. Second, we let the particle evolve until it reaches the position $x=a$. The time when $a$ is passed for the first time is recorded and the measurement is then repeated.
Above we have assumed that $y$ is a deterministic number. Of course we can consider also the random initial condition. In the latter case $y$ is drawn from a certain PDF $p_{0}(y)$, e.g. from the Gaussian PDF corresponding to initial equilibration in an optical trap \cite{Berg-SorensenRSI04}.

Finally, notice that the absorbing boundary at $x=a$ in fact stops rapidly diverging trajectories in the unstable potential in both the numerical simulations and the experiment. Statistics of first-passage times therefore allows more meaningful description of a stochastic motion for much longer times in contrast to frequently used time-dependent moments of position \cite{RadimPavelI}.

%%%%%%%%%%%%%%%%%%%%%%%%%%%%%%%%%%%%%%%%%%%%%%%%%%%%%%%%%%%%%%%%%%%%%%%%%%%%%%%%%%%%%%%%%%%%%%%%%%%%%%%%%%%
%%%%%%%%%%%%%%%%%%%%%%%%%%%%%%%%%%%%%%%%%%%%%%%%%%%%%%%%%%%%%%%%%%%%%%%%%%%%%%%%%%%%%%%%%%%%%%%%%%%%%%%%%%%
\section{First-passage time analysis of transient dynamics}
%%%%%%%%%%%%%%%%%%%%%%%%%%%%%%%%%%%%%%%%%%%%%%%%%%%%%%%%%%%%%%%%%%%%%%%%%%%%%%%%%%%%%%%%%%%%%%%%%%%%%%%%%%%
%%%%%%%%%%%%%%%%%%%%%%%%%%%%%%%%%%%%%%%%%%%%%%%%%%%%%%%%%%%%%%%%%%%%%%%%%%%%%%%%%%%%%%%%%%%%%%%%%%%%%%%%%%%

Let us denote by $f_{a}(t|y)$  the probability density of the first-passage time $\mathbf{T}(y\to a)$, i.e., $f_{a}(t|y) dt = {\rm Prob}\left\{ \mathbf{T}(y\to a) \in (t,t+dt) \right\}$. We illustrate this PDF for different starting point $y$ and the absorbing boundary position $a$ in Figs.~\ref{fig:fptpdf},~\ref{fig:fptasym}.  The first moment, i.e., the mean first-passage time (MFPT), is given by the integral
\begin{equation} 
\left<\mathbf{T}(y\to a) \right> = \int_{0}^{\infty}  dt\, t f_{a}(t|y) .
\end{equation}
The explicit expression of the MFPT for the specific $U(x)$ is derived by integration of the backward Fokker-Planck equation (\ref{BackwardFP}), using $f_a(t|y)=-\partial [\int_{a}^{\infty} dx\, p(x,t|y)] / \partial t$, see e.g.~Ref.~\cite{bookGardiner}. The result reads 
\begin{equation} 
\label{MFPT} 
 \left<\mathbf{T}(y\to a) \right> = \frac{1}{D}
\int_{a}^{y}d y_{1}  \int_{y_1 }^{\infty }d y_{2}\, {\rm e}^{-\beta\left[  U(y_{2}) - U(y_{1}) \right]}.
\end{equation}
We now introduce $\Delta \mathbf{T}(y\to a)= \mathbf{T}(y\to a)  -\left<\mathbf{T}(y\to a) \right>$, the fluctuation of the FPT around its mean value. The variance of the FPT can be derived along the same lines as MFPT (\ref{MFPT}), we obtain
\begin{equation} 
\label{VarianceofFPT}
\begin{split} 
\left<\left[\Delta \mathbf{T}(y\to a) \right]^{2} \right> = & 2
 \int_{a}^{y} d y_{1}  \int_{y_1 }^{\infty }d y_{2}\, {\rm e}^{-\beta\left[  U(y_{2}) - U(y_{1}) \right]} 
  \times \\
& \times \left[ \frac{\partial }{\partial y_2}
 \left< \mathbf{T}(y_{2}\to a)  \right>
 \right]^{2}. 
\end{split}
\end{equation}

The variance, or the standard deviation of FPT, is an important characteristics for thermally-induced transitions. Increase of the temperature $T$ may shorten MFPT, but simultaneously, the standard deviation of the FPT can be significantly increased. In the result, the passage time becomes shorter on average but simultaneously more uncertain. An interesting situation which occurs in the unstable potentials~(\ref{potentialU}) is that, by increasing $T$, we can achieve shorter MFPT $\left<\mathbf{T}(y\to a) \right>$ with the standard deviation $\sqrt{\left<\left[\Delta \mathbf{T}(y\to a) \right]^{2} \right>}$ decreasing comparably fast as $\left<\mathbf{T}(y\to a) \right>$. Before discussing this intriguing property of the nonlinear dynamics, let us review the linear case, where such effect cannot be achieved.

%%%%%%%%%%%%%%%%%%%%%%%%%%%%%%%%%%%%%%%%%%%%%%%%%%%%%%%%%%%%%%%%%%%%%%%%%%%%%%%%%%%%%%%%%%%%%%%%%%%%%%%%%%%
%%%%%%%%%%%%%%%%%%%%%%%%%%%%%%%%%%%%%%%%%%%%%%%%%%%%%%%%%%%%%%%%%%%%%%%%%%%%%%%%%%%%%%%%%%%%%%%%%%%%%%%%%%%
\subsection{The linear potential $U(x) = \mu x$}
%%%%%%%%%%%%%%%%%%%%%%%%%%%%%%%%%%%%%%%%%%%%%%%%%%%%%%%%%%%%%%%%%%%%%%%%%%%%%%%%%%%%%%%%%%%%%%%%%%%%%%%%%%%
%%%%%%%%%%%%%%%%%%%%%%%%%%%%%%%%%%%%%%%%%%%%%%%%%%%%%%%%%%%%%%%%%%%%%%%%%%%%%%%%%%%%%%%%%%%%%%%%%%%%%%%%%%%

For the linear potential 
\begin{equation}
U(x) =  \mu x,
\label{LinPot}
\end{equation} 
  the PDF of the particle position is Gaussian. If the particle departs at $t=0$ from $x=y$, then the mean value of its position  and the variance are equal to 
\begin{eqnarray}
\label{meanlinear} 
& &\left< x(t) \right> =y -  \frac{\mu}{\gamma} t, \\ 
\label{varlinear} 
& &\left< [\Delta x(t) ]^{2} \right> = 2 \frac{k_B T}{\gamma} t, 
\end{eqnarray} 
respectively. 
The first equation tells us that the overdamped particle moves with a constant average velocity $v$, which equals to the potential strength divided by the friction coefficient, 
$v=- {\mu }/{\gamma}$. 
The second equation, Eq.~(\ref{varlinear}), quantifies the effect of the thermal noise: the variance $\left< \left[ \Delta x(t) \right]^{2} \right>$  grows linearly with time.
The latter fact should be compared to the corresponding result for the fluctuations of the FPT~(\ref{varTlin}) as we discuss below.

%%%%%%%%%%%%%%%%%%%%%%%%%%%%%%%%%%%%%%%%%%%%%%%%%%%%%%%%%%%%%%%%%%%%%%%%%%%%%%%%%%%%%%%%%%%%%%%%%%%%%%%%%%%
\subsubsection{Mean first-passage times}
%%%%%%%%%%%%%%%%%%%%%%%%%%%%%%%%%%%%%%%%%%%%%%%%%%%%%%%%%%%%%%%%%%%%%%%%%%%%%%%%%%%%%%%%%%%%%%%%%%%%%%%%%%%

The complementary information about the dynamics of the particle can be inferred from the distribution of the first-passage time $\mathbf{T}(y\to a)$. Analytical expression for the MFPT is obtained from  Eq.~(\ref{MFPT}). We assume that $y>0$ and $a<0$ as illustrated in Fig.~\ref{fig:schematic}, then for the linear potential (\ref{LinPot}) the result reads
\begin{equation} 
\label{MFPTlinV0} 
 \left<\mathbf{T}(y\to a) \right> = \frac{\gamma}{\mu }\left(y-a\right).
\end{equation} 
Eq.~(\ref{MFPTlinV0}) reflects the very same basic fact as Eq.~(\ref{meanlinear}) does. Namely, that the Brownian particle moves with a constant mean velocity $v=- {\mu }/{\gamma}$, independently of the temperature $T$. We can reshape it into the form rather similar  to Eq.~(\ref{meanlinear}), we have $
a=y - \mu \left<\mathbf{T}(y\to a) \right>/\gamma$  (to see the similarity replace $a\to x(t)$, $\left<\mathbf{T}(y\to a) \right>\to t$).

There are two inherent features of the MFPT which will be completely different in the nonlinear case as compared to the present linear one.  
The first striking consequence of the linearity of the potential is that the first moment of FPT~(\ref{MFPTlinV0}) is \emph{independent of the noise strength} (i.e.\ of the temperature $T$), cf.~Eq.~(\ref{MFPTxnytoa}) for the nonlinear case, where the power-law dependence on $T$ occurs. 

Second, notice that the relative time to exit the positive half-line $x>0$ for the first time is proportional to the initial distance from the origin, 
\begin{equation} 
\frac{\left< {\bf T} (y \to 0) \right>}{\left< {\bf T} (y \to a) \right>}
= \frac{y}{y-a},
\quad  
\frac{\left< {\bf T} (0 \to a) \right>}{\left< {\bf T} (y \to a) \right>}
= \frac{-a}{y-a}. 
\end{equation}
In particular, for the symmetric situation when $a=-y$,  FPTs ${\bf T} (y \to 0)$, ${\bf T} (0 \to -y)$ {\em are identically distributed} and the above ratios of MFPTs are equal to $1/2$. This is no longer true for the nonlinear potentials, where FPTs ${\bf T} (y \to 0)$, ${\bf T} (0 \to -y)$, differ in distribution and the corresponding ratios are determined by the exponent $n$, see Eqs.~(\ref{ratioMFPTs3}),~(\ref{ratioMFPTs5}).

%%%%%%%%%%%%%%%%%%%%%%%%%%%%%%%%%%%%%%%%%%%%%%%%%%%%%%%%%%%%%%%%%%%%%%%%%%%%%%%%%%%%%%%%%%%%%%%%%%%%%%%%%%%
\subsubsection{Signal-to-noise ratio}
%%%%%%%%%%%%%%%%%%%%%%%%%%%%%%%%%%%%%%%%%%%%%%%%%%%%%%%%%%%%%%%%%%%%%%%%%%%%%%%%%%%%%%%%%%%%%%%%%%%%%%%%%%%

The thermal noise presented in the model manifests itself through the dispersion of the particle position around the mean value. This dispersion, \emph{at a given time}, is quantified by Eq.~(\ref{varlinear}). Alternatively, the strength of the thermal noise could be quantified using the dispersion of the FPT. For the linear potential, Eq.~(\ref{VarianceofFPT}) yields
\begin{equation} 
\label{varTlin}
\left<[ \Delta \mathbf{T}(y\to a)]^{2} \right> = \frac{2 k_B T}{\gamma} \left( \frac{\gamma}{\mu } \right)^{3} \left( y-a\right). 
\end{equation} 
Hence the variance of the FPT grows linearly with the distance $(y-a)$ from the initial position. This should be compared with Eq.~(\ref{varlinear}) where, on the other hand, the variance of the position grows linearly with time.  
We can also rewrite the last result in the form similar to the fluctuation-dissipation theorem from the linear-response theory \cite{JarzynsiFR}. In fact, Eq.~(\ref{varTlin}) relates the variance of the FPT to its mean multiplied by the temperature:
\begin{equation} 
\left<[ \Delta \mathbf{T}(y\to a)]^{2} \right> = \frac{2k_{B} T}{\gamma v^{2}}  \left<\mathbf{T}(y\to a) \right>.
\end{equation} 

The first definition of the signal-to-noise ratio that we use in the following reads
\begin{equation}
\label{SNRlin1}
{\rm SNR} = \frac{ \left<\mathbf{T}(y\to a) \right>
 }{\sqrt{\left<[ \Delta \mathbf{T}(y\to a)]^{2} \right>}} =
\sqrt{ \frac{1}{2} \frac{\mu (y-a)}{k_B T}}.
\end{equation} 
It measures the relative significance of the mean value of the FPT as compared to its fluctuations. On the right-hand side of Eq.~(\ref{SNRlin1}) we recognize  the potential difference between the initial position and the position of the boundary divided by the thermal energy. This means that, for the linear potential, the SNR decreases with temperature as the variance of the FPT ${\bf T}(y\to a)$ grows. Increasing the temperature has purely negative effect in the linear case, it is therefore better to avoid it. The situation will be different in the nonlinear case, where the potential induces completely different temperature dependencies. 

We have described thermally-induced effects present for the linear potential \eqref{LinPot} to clearly discriminate qualitatively different effects caused by nonlinear potentials described in the next Section.   
Despite the fact that the linear potential \eqref{LinPot} yields a rather simple noise-induced behavior, its modifications are frequently used to model more complex situations. Variety of analytically tractable models based on time-dependent or/and piece-wise linear potentials can be found in Refs.~\cite{Malakhov1991I, Malakhov1991II, Agudov1993, AgudovSpagnolo2001,  Chvosta2003, DubkovPRE2004, SpagnoloEPJB2004, DubkovAgudov2004ActaB, Dubkov2004ActaB, RyabovChvosta2011, UrdapilletaPRE2011, Urdapilleta2011, Urdapilleta2012, Metzler2012}.

%%%%%%%%%%%%%%%%%%%%%%%%%%%%%%%%%%%%%%%%%%%%%%%%%%%%%%%%%%%%%%%%%%%%%%%%%%%%%%%%%%%%%%%%%%%%%%%%%%%%%%%%%%%
%%%%%%%%%%%%%%%%%%%%%%%%%%%%%%%%%%%%%%%%%%%%%%%%%%%%%%%%%%%%%%%%%%%%%%%%%%%%%%%%%%%%%%%%%%%%%%%%%%%%%%%%%%%
\subsection{The nonlinear potential $U(x) = \mu x^n$}
\label{subsec:nonlinear}
%%%%%%%%%%%%%%%%%%%%%%%%%%%%%%%%%%%%%%%%%%%%%%%%%%%%%%%%%%%%%%%%%%%%%%%%%%%%%%%%%%%%%%%%%%%%%%%%%%%%%%%%%%%
%%%%%%%%%%%%%%%%%%%%%%%%%%%%%%%%%%%%%%%%%%%%%%%%%%%%%%%%%%%%%%%%%%%%%%%%%%%%%%%%%%%%%%%%%%%%%%%%%%%%%%%%%%%

The first-passage properties of the particle diffusing in the nonlinear potential 
\begin{equation} 
\label{nonlinearpot}
U(x) = \mu x^n, \qquad n=3, 5, \ldots
\end{equation} 
are considerably different from the linear case $n=1$. Our primary interest is in the cubic potential $n=3$, yet the following analysis will be performed for a general odd $n$.

An intriguing property of the nonlinear potential (\ref{nonlinearpot}), which is absent in the linear case, is that the Brownian particle {\em can travel an infinite distance in a finite time}. 
Formally, this can be observed if we perform limits $y\to \infty$ and/or $a\to -\infty$ in the expression for the MFPT~(\ref{MFPT}). When $n=1$, the limits yield an infinite value, however, for $n\geq 3$ the result is finite, e.g.\ $\left< {\bf T}(\infty \to -\infty) \right><\infty$. 
Physically we should interpret the above limits as follows. To this end we introduce the length scale $l$ determined by the strength of thermal fluctuations as compared to value of the nonlinear potential $U(x)$:  
\begin{equation}
\label{ldef} 
U(x)/k_B T=\left(x/l\right)^{n},  
\quad 
\mbox{giving}
\quad  
l=\left(k_B T/\mu  \right)^{1/n}.
\end{equation} 
Within the interval $(-l,l)$ the potential $U(x)$ is smaller or comparable to the thermal energy $k_B T$, thus in $(-l,l)$ the dynamics is significantly influenced by thermal fluctuations. 
 When the particle leaves the interval $(-l,l)$, the potential $U(x)$ prevails over the thermal energy  since the (absolute) value of the potential rapidly increases with increasing (absolute) value of $x$. Thus the dynamics of a Brownian particle away from  $(-l,l)$ is much faster than that in $(-l,l)$ and, at the same time, the noise has a negligible influence on it. This behavior is clearly illustrated on simulated trajectories in Fig.~\ref{fig:schematic}. 

From the above reasoning it follows that if we place the initial position $y$ and the absorbing boundary $a$ away from $(-l,l)$, i.e., if 
\begin{equation} 
\label{bounds} 
a \ll - \left( k_B T / \mu \right)^{1/n},\quad {\rm and} \quad
y \gg \left( k_B T / \mu \right)^{1/n}, 
\end{equation} 
then we may approximate the FPT ${\bf T}(a\to y)$ by ${\bf T}(\infty \to -\infty)$. The approximation actually yields a compact expression for the MFPT and reveals its temperature dependence.  

Notice that the length scale $l$ varies differently with the power of the nonlinearity $n$ depending on whether $k_B T \ll \mu$ or $k_B T  \gg \mu$. In the low-temperature limit, $k_B T \ll \mu$, the interval $(-l,l)$, where the dynamics is influenced by the noise, becomes larger as $n$ increases. On the other hand, in the weak potential, $k_B T  \gg \mu$, the length $l$ decreases with $n$.

%%%%%%%%%%%%%%%%%%%%%%%%%%%%%%%%%%%%%%%%%%%%%%%%%%%%%%%%%%%%%%%%%%%%%%%%%%%%%%%%%%%%%%%%%%%%%%%%%%%%%%%%%%%
\subsubsection{Mean first-passage times}
%%%%%%%%%%%%%%%%%%%%%%%%%%%%%%%%%%%%%%%%%%%%%%%%%%%%%%%%%%%%%%%%%%%%%%%%%%%%%%%%%%%%%%%%%%%%%%%%%%%%%%%%%%%

The FPT is additive in one dimension, thus we have 
\begin{equation} 
\label{MFPTya} 
\left<  {\bf T}(y\to a) \right> =  \left< {\bf T}(y\to 0)  \right> 
 +  \left< {\bf T}(0\to a)  \right>.
\end{equation} 
Let us now approximate the both terms on the right-hand side assuming that inequalities (\ref{bounds}) are satisfied. The result is given in Eq.~(\ref{MFPTxnytoa}).  

The MFPT $\left< {\bf T} (y\to 0)  \right>$ to the origin for the particle that starts from a large distance $y$,  $y \gg \left( k_B T / \mu \right)^{1/n}$, can be decomposed into the temperature-dependent and the temperature-independent parts with a rather clear physical meaning. In accordance with the additivity of FPTs~(\ref{MFPTya}) we can write  
\begin{equation} 
\label{Tytozero}
\left< {\bf T}(y\to 0)\right>= \left< {\bf T}(\infty \to 0)\right> -  \left< {\bf T}(\infty \to y)\right>,
\end{equation}
where the MFPT $\left< {\bf T} (\infty \to 0) \right>$ is given by the product of the power of temperature and the temperature-independent constant determined by the parameters of the nonlinear potential. From the general result (\ref{MFPT}) we obtain 
\begin{equation} 
\begin{split}
\left< {\bf T} (\infty \to 0) \right> = & \frac{\gamma }{\mu} \left( \frac{\mu }{k_B T} \right)^{1-2/n}
  \left[ \Gamma\!\left(1+\frac{1}{n} \right)\right]^{2} \\
 & \times  \frac{1}{2\sin\!\left( \frac{\pi}{2} (1-\frac{2}{n}) \right)} . 
\end{split} 
\label{MFPTinftyto0}
\end{equation} 
The MFPT~(\ref{MFPTinftyto0}) accounts for the dynamics on the whole half-line $(0,\infty)$ including the neighborhood of the origin $(0,l)$, where the thermal noise has a profound impact on the particle motion. Contrary to that, the second term on the right-hand side of Eq.~(\ref{Tytozero}), $\left< {\bf T} (\infty  \to y)  \right>$, corresponds to fast passage through high values of the potential $U(x)$, where the thermal noise can be neglected. We thus obtain the approximate expression for MFPT $\left< {\bf T} (y  \to 0)  \right>$ as the sum of the temperature-dependent and temperature-independent terms:
\begin{equation} 
\label{MFPTinfty0} 
\left< {\bf T} (y\to 0)  \right> \approx 
\left< {\bf T} (\infty \to 0 )  \right> - 
\frac{\gamma }{\mu n (n-2)y^{n-2}}.
\end{equation}
The approximate expression  $\left< {\bf T} (\infty  \to y)  \right> \approx  \left. \gamma \right/ \left[\mu n (n-2)y^{n-2} \right]$  can be derived either from the general result (\ref{MFPT}) or directly by integration of the Langevin equation (\ref{Langevin}) without the noise term ($D=0$), cf.~Eq.~(\ref{tauinfty}).

%%%%%%%%%%%%%%%%%%%%%%%%%%%%%%%%%%%%%%%%%%%%%%%%%%%%%%%%%%%%%%%%%%%%%%%%%%%%%%%%%%%%%%%%%%%%%%%%%%%%%%%%%%% 
\begin{figure}[t!]
\includegraphics[scale=0.81]{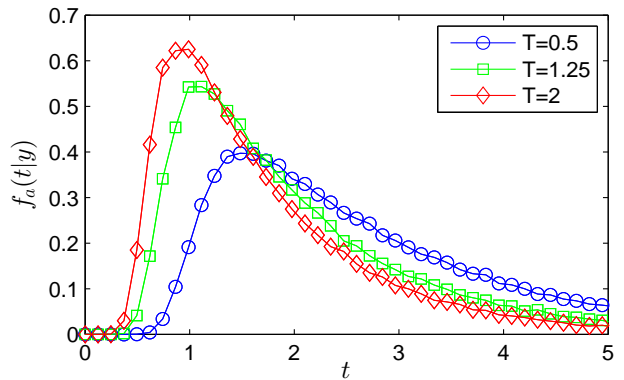}
\caption{\label{fig:fptpdf} 
The simulated PDF $f_{a}(t|y)$ of the first-passage time ${\bf T}(y\to a)$ for the cubic potential and three temperatures. Parameters used: $y=4$, $a=-4$, $\gamma=k_B=\mu=1$. Trajectories of the Brownian motion were obtained by a numerical integration of the Langevin equation~(\ref{Langevin}) using the Euler-Maruyama method \cite{bookKloeden}. Each histogram of FPTs was computed from $10^{6}$ simulated trajectories ($N_{traj}=10^{6}$).} 
\end{figure}
%%%%%%%%%%%%%%%%%%%%%%%%%%%%%%%%%%%%%%%%%%%%%%%%%%%%%%%%%%%%%%%%%%%%%%%%%%%%%%%%%%%%%%%%%%%%%%%%%%%%%%%%%%%

Reasoning along the similar lines yields the asymptotic approximation for the MFPT 
$\left< {\bf T} (0\to a)  \right> $
when the particle starts at the origin and the absorbing boundary is far enough from the origin, i.e., $a \ll -1\left( k_B T / \mu \right)^{1/n}$. Again we decompose the result into two terms 
\begin{equation} 
\label{MFPT0minfty}
\left< {\bf T} (0\to a)  \right> \approx 
\left< {\bf T} (0\to -\infty )  \right> - 
\frac{\gamma }{\mu n (n-2)|a|^{n-2}},
\end{equation} 
where 
\begin{equation}
\begin{split}
\left< {\bf T} (0 \to -\infty) \right> = & \frac{\gamma }{\mu} \left( \frac{\mu }{k_B T} \right)^{1-2/n}
  \left[ \Gamma\!\left(1+\frac{1}{n} \right)\right]^{2} \\
 & \times \left(1+ \frac{1}{2\sin\!\left( \frac{\pi}{2} (1-\frac{2}{n}) \right)} \right), 
\end{split} 
\end{equation}
and the  temperature-independent term is nothing but the approximation of the MFPT from $a$ to minus infinity, 
$\left< {\bf T} (a\to -\infty)  \right> \approx  \left. \gamma \right/ \left[\mu n (n-2)|a|^{n-2} \right]$.

%%%%%%%%%%%%%%%%%%%%%%%%%%%%%%%%%%%%%%%%%%%%%%%%%%%%%%%%%%%%%%%%%%%%%%%%%%%%%%%%%%%%%%%%%%%%%%%%%%%%%%%%%%% 
\begin{figure}[t!]
\includegraphics[scale=0.81]{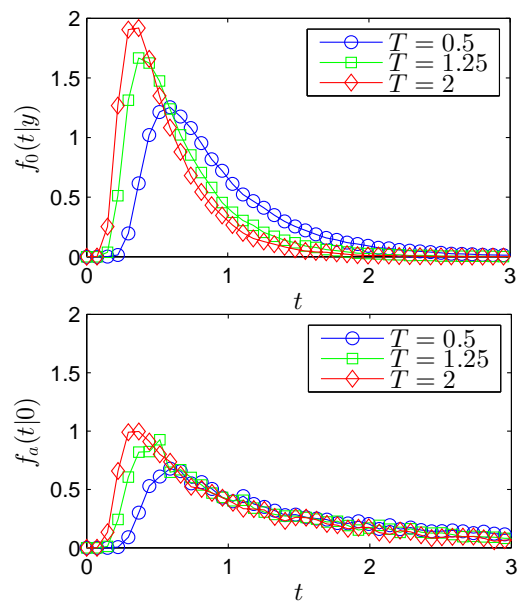}
\caption{\label{fig:fptasym} 
The simulated PDF $f_{0}(t|y)$ of the FPT ${\bf T}(y\to 0)$ (the upper panel) and the simulated PDF $f_{a}(t|0)$ of the FPT ${\bf T}(0\to a)$ for the cubic potential and three temperatures. Parameters used: $y=4$, $a=-4$, $\gamma=k_B=\mu=1$, $N_{traj}=10^{6}$ .} 
\end{figure} 
%%%%%%%%%%%%%%%%%%%%%%%%%%%%%%%%%%%%%%%%%%%%%%%%%%%%%%%%%%%%%%%%%%%%%%%%%%%%%%%%%%%%%%%%%%%%%%%%%%%%%%%%%%%

Summing up the two results (\ref{MFPT0minfty}), (\ref{MFPTinfty0}), we obtain the approximation of the MFPT from Eq.~(\ref{MFPTya}),
\begin{equation} 
\label{MFPTxnytoa} 
\begin{split} 
\left< {\bf T} (y\to a) \right> \approx & 
\left< {\bf T} (\infty \to -\infty) \right> \\
& - 
\frac{\gamma }{\mu n (n-2)} 
\left( \frac{1}{y^{n-2}}+ \frac{1}{|a|^{n-2}}\right), 
\end{split} 
\end{equation} 
where the MFPT from $y=\infty$ to $a=-\infty$, is given by
\begin{equation} 
\label{MFPTinfinf} 
\begin{split} 
\left< {\bf T} (\infty \to -\infty) \right> = & \frac{\gamma }{\mu} \left( \frac{\mu }{k_B T} \right)^{1-2/n}
  \left[ \Gamma\!\left(1+\frac{1}{n} \right)\right]^{2} \\
 & \times \left(1+ \frac{1}{ \sin\!\left( \frac{\pi}{2} (1-\frac{2}{n}) \right)} \right).  
\end{split} 
\end{equation} 
Note that, for a particular case $n=3$, the formula \eqref{MFPTinfinf} was first derived in Ref.~\cite{horsthemke89}, see also Refs.~\cite{Lindner03, Sancho91,  ReimannPRE02, ReimannBroeck94} for its use in various contexts.  
In the limit $T\to 0$ the MFPT $\left< {\bf T} (\infty \to -\infty) \right>$ diverges since the origin $x=0$ is the (unstable) equilibrium point which cannot be crossed without the aid of the thermal noise. Fig.~\ref{fig:fptpdf} demonstrates that the mean of the FPT PDF grows with decreasing $T$ and, at the same time, the tail becomes heavier. 
Increasing the  temperature $T$ has therefore a positive effect on the passage time, i.e., the MFPT becomes smaller. 
Temperature dependence is the same for $\left< {\bf T} (\infty \to -\infty) \right>$, $\left< {\bf T} (0 \to -\infty) \right>$ and $\left< {\bf T} (\infty \to 0) \right>$ and therefore, it is both simple and inherent temperature-based characteristics of the unstable potential.

%%%%%%%%%%%%%%%%%%%%%%%%%%%%%%%%%%%%%%%%%%%%%%%%%%%%%%%%%%%%%%%%%%%%%%%%%%%%%%%%%%%%%%%%%%%%%%%%%%%%%%%%%%% 
\begin{figure}[t!]
\includegraphics[scale=0.75]{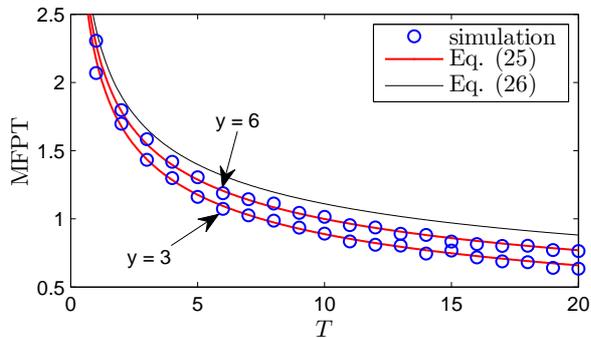}
\caption{\label{fig:mfpt} MFPT $\left< {\bf T}(y\to a) \right>$ as the function of the temperature $T$ computed from $1000$ simulated trajectories. Parameters used: $a=-y$, $\gamma=k_B=\mu=1$. } 
\end{figure} 
%%%%%%%%%%%%%%%%%%%%%%%%%%%%%%%%%%%%%%%%%%%%%%%%%%%%%%%%%%%%%%%%%%%%%%%%%%%%%%%%%%%%%%%%%%%%%%%%%%%%%%%%%%%

It turns out that the formula~(\ref{MFPTxnytoa}) describes the MFPT surprisingly well even for relatively small values of $a$ and $y$. The MFPT computed from a {\em small} sample of simulated trajectories is depicted in Fig.~\ref{fig:mfpt} for $n=3$ and for different values of $a$ and $y$. We have intentionally chosen the small sample of trajectories in order to verify that the MFPT~(\ref{MFPTxnytoa}) can be reliably measured in the actual experiment. Fig.~\ref{fig:mfpt} also demonstrates that the limiting formula~(\ref{MFPTinfinf}), despite the fact that it dictates the temperature-dependence of the MFPT, dictates {\em the theoretical upper bound} on MFPT for a given temperature. 

\begin{figure}[t!]
\includegraphics[scale=0.81]{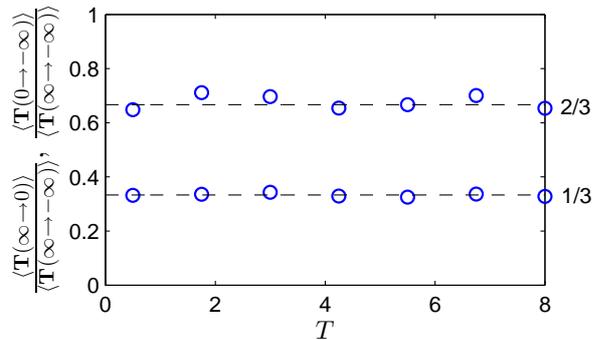}
\caption{\label{fig:ratios}  Even for small number of trajectories, the fractions~(\ref{ratioMFPTs3}) are obtained in simulation with a sufficient precision. We have used $n=3$, $N_{traj}=1000$, $y=5$, $a=-5$,  $dt=10^{-4}$. } 
\end{figure}

Presently, in contrast to the linear case, FPTs ${\bf T} (y \to 0)$, ${\bf T} (0 \to -y)$ are no longer identically distributed. The asymmetry of the FPT PDFs is demonstrated in Fig.~\ref{fig:fptasym}. In the limit $y\to \infty$, the asymmetry can be expressed by appealing formulas which involve ratios of MFPTs. For the cubic potential ($n=3$) we obtain the fractions  
\begin{equation} 
\label{ratioMFPTs3} 
\frac{\left< {\bf T} (0 \to -\infty) \right>}{\left< {\bf T} (\infty \to -\infty) \right>}
= \frac{2}{3}, 
\quad 
\frac{\left< {\bf T} (\infty \to 0) \right>}{\left< {\bf T} (\infty \to -\infty) \right>} = \frac{1}{3}, 
\end{equation}
while for $n=5$ the fractions read  
\begin{equation} 
\label{ratioMFPTs5} 
\begin{split}
& \frac{\left< {\bf T} (0 \to -\infty) \right>}{\left< {\bf T} (\infty \to -\infty) \right>}
= \frac{1}{2}\left(1+\frac{1}{\sqrt{5}} \right),
\\
& \frac{\left< {\bf T} (\infty \to 0) \right>}{\left< {\bf T} (\infty \to -\infty) \right>}
= \frac{1}{2}\left(1-\frac{1}{\sqrt{5}} \right).
\end{split}
\end{equation}
The fractions are not only temperature-independent, but also they do not depend on the strength $\mu$ of nonlinearity and damping constant $\gamma$. Although the overall MFPT is shortened for higher temperatures $T$, a departure time from the origin becomes relatively longer. Simultaneously, the arrival time is much shorter for higher order $n$ of the nonlinearity. The analytical results for fractions can be quickly approached for small number of trajectories, as is visible from stochastic simulations depicted in Fig.~\ref{fig:ratios}. The fractions can be therefore easily measured in the actual experiment. Their second advantage is that they allow to test the nonlinearity of the potential without fast measurement of details of quickly diverging trajectories.

%%%%%%%%%%%%%%%%%%%%%%%%%%%%%%%%%%%%%%%%%%%%%%%%%%%%%%%%%%%%%%%%%%%%%%%%%%%%%%%%%%%%%%%%%%%%%%%%%%%%%%%%%%% 
%%%%%%%%%%%%%%%%%%%%%%%%%%%%%%%%%%%%%%%%%%%%%%%%%%%%%%%%%%%%%%%%%%%%%%%%%%%%%%%%%%%%%%%%%%%%%%%%%%%%%%%%%%%

%%%%%%%%%%%%%%%%%%%%%%%%%%%%%%%%%%%%%%%%%%%%%%%%%%%%%%%%%%%%%%%%%%%%%%%%%%%%%%%%%%%%%%%%%%%%%%%%%%%%%%%%%%% 
\begin{figure}[t!]
\includegraphics[scale=0.75]{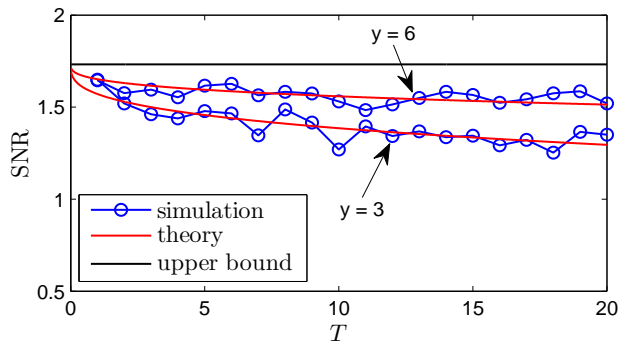}
\caption{\label{fig:snr} SNR as the function of the temperature $T$ computed from $1000$ simulated trajectories for $n=3$ and the same parameters as used in Fig.~\ref{fig:mfpt}. The upper bound on SNR is given by the temperature-independent result (\ref{upperbound}). The theoretical curves (red solid lines) are computed as $  \left<\mathbf{T}(y \to a) \right> / \sqrt{ \left<[ \Delta \mathbf{T}(\infty \to -\infty )]^{2} \right> }$, where the first moment of the FPT $\left<\mathbf{T}(y \to a) \right>$ is given by Eq.~(\ref{MFPTxnytoa}) and its asymptotic standard deviation by Eq.~(\ref{2ndMomentSigeti}).} 
\end{figure} 
%%%%%%%%%%%%%%%%%%%%%%%%%%%%%%%%%%%%%%%%%%%%%%%%%%%%%%%%%%%%%%%%%%%%%%%%%%%%%%%%%%%%%%%%%%%%%%%%%%%%%%%%%%%

%%%%%%%%%%%%%%%%%%%%%%%%%%%%%%%%%%%%%%%%%%%%%%%%%%%%%%%%%%%%%%%%%%%%%%%%%%%%%%%%%%%%%%%%%%%%%%%%%%%%%%%%%%%
\subsubsection{Signal-to-noise ratio}
%%%%%%%%%%%%%%%%%%%%%%%%%%%%%%%%%%%%%%%%%%%%%%%%%%%%%%%%%%%%%%%%%%%%%%%%%%%%%%%%%%%%%%%%%%%%%%%%%%%%%%%%%%%
For the linear potential the variance of the FPT is proportional to the temperature, see Eq.~(\ref{varTlin}). For the nonlinear potential the temperature-dependence of the variance is controlled by the exponent $n$.  When the initial position $y$ and the position of the boundary $a$ are far enough from the origin (i.e., when they satisfy inequalities (\ref{bounds})), the variance moment of the FPT~(\ref{VarianceofFPT}) becomes 
\begin{equation} 
\label{2ndMomentInfInf} 
 \left<\left[\Delta \mathbf{T}(\infty \to -\infty)\right]^{2} \right> = 
 \left( \frac{\gamma }{\mu} \right)^{2} \left( \frac{\mu }{k_B T} \right)^{2(1-2/n)}  A(n), 
\end{equation} 
where the factor $A(n)$ is expressed by the four integrals from Eq.~(\ref{VarianceofFPT}) and it depends solely on $n$. In particular, for $n=3$, the exact expression for the variance was obtained by Sigeti and Horsthemke \cite{horsthemke89}. Their remarkable formula reads 
\begin{equation} 
\label{2ndMomentSigeti} 
 \left<\left[\Delta \mathbf{T}(\infty \to -\infty)\right]^{2} \right> = \frac{1}{3}
 \left( \frac{\gamma }{\mu} \right)^{2} \left( \frac{\mu }{k_B T} \right)^{2/3}, 
 \quad n=3.
\end{equation}  
This implies that the variance of the FPT {\em decreases} with temperature. Similar conclusion holds true also for $\left<\left[\Delta \mathbf{T}(0 \to -\infty)\right]^{2} \right> $ and $\left<\left[\Delta \mathbf{T}(\infty \to 0)\right]^{2} \right> $. These facts can be also qualitatively deduced from the simulated histograms of FPT densities, cf.~the tails of FPT PDFs in Figs.~\ref{fig:fptpdf} and~\ref{fig:fptasym}. 
 
When the potential is linear, the signal-to-noise ratio (\ref{SNRlin1})) decreases with $T$. Presently, however, the situation is rather different. The first signal-to-noise ratio, defined similarly to (\ref{SNRlin1}), 
\begin{equation} 
\label{SNRt}
{\rm SNR} = \frac{\left<\mathbf{T}(\infty \to -\infty) \right>
 }{ \sqrt{ \left<[ \Delta \mathbf{T}(\infty \to -\infty )]^{2} \right> }} , 
\end{equation} 
{\em does not depend} on the temperature $T$. It happens because both the square of the mean (\ref{MFPTinfinf}) and the second moment (\ref{2ndMomentInfInf}) are proportional to the same power of $T$.  In particular, for the cubic potential we get
\begin{equation}
\label{upperbound}
{\rm SNR} = \sqrt{3}, \quad n=3.
\end{equation}
For higher values of $n$ the exact result is not known and~(\ref{SNRt}) should be evaluated numerically or by a scaling approach as in Ref.~\cite{Vilar19991}. 

Finally, it is rather instructive to deduce the temperature dependence of ${\bf T}(\infty\to -\infty)$  from {\em simple scaling arguments}. When inequalities (\ref{bounds}) hold, both the length scale associated with the initial position and the length scale associated with the position of the absorbing boundary disappears from the problem. Hence all the results should depend just on the characteristic length scale $l$ which stems from the ratio of potential and thermal energy~\eqref{ldef}: $U(x)/k_B T=(x/l)^{n}$,  $l= \left( k_B T/\mu \right)^{1/n}$. At the same time, the Brownian scaling relates the length scale to the characteristic time $\tau$ as follows, $l^{2} \sim ( k_{B}T/\gamma )\tau $. Hence the characteristic time $\tau$ in the problem (which in our case fixes the scale of the FPT) must depend on the temperature as 
\begin{equation} 
\label{scalingtau}
\tau \sim \frac{\gamma}{\mu} \left( \frac{\mu }{k_B T}\right)^{1-2/n},
\end{equation}
which correctly predicts the temperature-dependence of all moments of the FPT discussed above. 

In the limit of the extremely steep potential $n\to \infty$, $l\to 1$ and from Eq.~(\ref{scalingtau}) we have $ D \tau \sim  1$. For a finite $n$, the time scale $\tau$ behaves differently for $k_B T \ll \mu$ and $k_B T \gg \mu$. When thermal fluctuations are small, $k_B T \ll \mu$, the FPT increases with growing $n$. On the other hand, when thermal energy prevails upon potential strength, $k_B T \gg \mu$, the FPT decreases with $n$.

%%%%%%%%%%%%%%%%%%%%%%%%%%%%%%%%%%%%%%%%%%%%%%%%%%%%%%%%%%%%%%%%%%%%%%%%%%%%%%%%%%%%%%%%%%%%%%%%%%%%%%%%%%%
%%%%%%%%%%%%%%%%%%%%%%%%%%%%%%%%%%%%%%%%%%%%%%%%%%%%%%%%%%%%%%%%%%%%%%%%%%%%%%%%%%%%%%%%%%%%%%%%%%%%%%%%%%%
\section{Non-equilibrium steady state in unstable potential and experimental outlook} 
%%%%%%%%%%%%%%%%%%%%%%%%%%%%%%%%%%%%%%%%%%%%%%%%%%%%%%%%%%%%%%%%%%%%%%%%%%%%%%%%%%%%%%%%%%%%%%%%%%%%%%%%%%% 
%%%%%%%%%%%%%%%%%%%%%%%%%%%%%%%%%%%%%%%%%%%%%%%%%%%%%%%%%%%%%%%%%%%%%%%%%%%%%%%%%%%%%%%%%%%%%%%%%%%%%%%%%%%

Experimental observation of stochastic trajectories in unstable potentials is a challenging task. Above, we have proposed how to characterize and measure {\em transient} dynamics of such trajectories using their first-passage properties. 
Let us now describe, how we can measure also {\em stationary} long-time distribution of such trajectories.

A possible experiment runs as follows, cf.~Fig.~(\ref{fig:schematic}). At the initial time the particle is localized at the initial position $y$ in a potential well such as $U(x)=\mu x^{2}$. It is formed by a focused laser beam (referred to as optical tweezers \cite{Jonesbook15}) and the beam center is situated at the initial position $y$. Then this trapping potential is switched off and substituted by the nonlinear potential of the investigated shape $U(x)=\mu x^{n}$ with odd $n$ which pushes the particle towards $x=-\infty$. The particle trajectories are followed by an optical microscope, recorded by a CCD.  
As soon as the particle reaches the absorbing boundary at $a$, the nonlinear potential is switched off and substituted by the trapping potential (optical tweezers) and the particle is returned to its initial position $y$ via repositioning the laser beam focus and the experiment can be repeated. However, optical forces are generally non-conservative in three-dimensions, therefore the counter-propagating geometries should be considered to get conservative optical potentials \cite{RadimPavelII, DivittOL15} following the spatial profile of the optical intensity of the laser beam. Spatial light modulator provides fast dynamic modifications of the spatial profile of the laser beam optical intensity to get the desired optical potential.
	
In constant confining potentials such as $U(x)=\mu x^{2}$, the PDF for the particle position converges towards the Gibbs canonical distribution $p_{\rm eq}(x)$ at long times. This fact, for example, can be exploited in order to determine actual potential which acts on the particle trapped in optical tweezers \cite{Speck06}, since $U(x) \sim - k_B T \log(p_{\rm eq }(x))$ in this case.  
Contrary to this, the potential $U(x)=\mu x^{n}$ with odd $n$ pushes the particle towards $x=-\infty$, which implies that no non-trivial long-time limit of the solution of the Fokker-Planck equation exists. However, by the micromanipulation with the particle in optical tweezers, we can create a \emph{non-equilibrium steady state}, where the stationary PDF $p_{\rm st}(x)$ (different from the Gibbs canonical distribution) will reflect the type of the nonlinearity of the unstable potential. Moreover, the order of the nonlinearity $n$ manifests itself through the position of the maximum of $p_{\rm st}(x)$, see Eq.~(\ref{positionofmaximum}), which is an easily measurable quantity. The maximum of the PDF can be approximately determined from a small sample of trajectories. Its shift with temperature bears all essential information about the underlying nonlinear potential.

%%%%%%%%%%%%%%%%%%%%%%%%%%%%%%%%%%%%%%%%%%%%%%%%%%%%%%%%%%%%%%%%%%%%%%%%%%%%%%%%%%%%%%%%%%%%%%%%%%%%%%%%%%%
\begin{figure}[t!]
\includegraphics[scale=0.8]{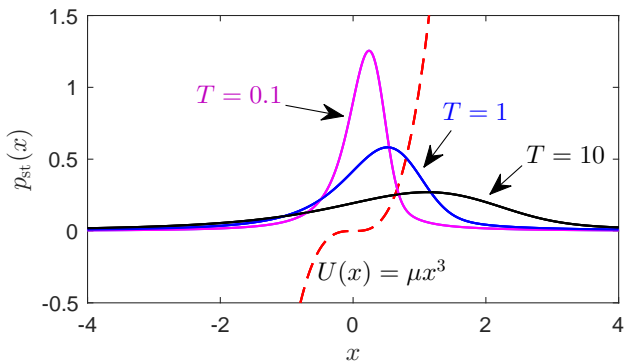}
\caption{\label{fig:PDFstac} 
Stationary PDF (\ref{PDFstac}) for three temperatures. Parameters used: $k_B=\gamma=\mu=1$. } 
\end{figure}
%%%%%%%%%%%%%%%%%%%%%%%%%%%%%%%%%%%%%%%%%%%%%%%%%%%%%%%%%%%%%%%%%%%%%%%%%%%%%%%%%%%%%%%%%%%%%%%%%%%%%%%%%%%

After several repeats of the experiment (many cycles), the PDF for the position approaches the limiting value 
\begin{equation}  
p_{\rm st}(x)=\lim_{t\to \infty}p(x,t|y),
\end{equation}
which describes the PDF profile in the non-equilibrium steady state. Obviously, now the PDF $p_{\rm st}(x)$ is not given by the Gibbs canonical distribution. Yet, still we can use the PDF $p_{\rm st}(x)$ to probe the actual potential (force) acting on the particle. 

Let us now treat the simplest case. We assume that the inequalities (\ref{bounds}) are satisfied. The stationary PDF $p_{\rm st}(x)$ follows from the fact that in the steady state the  probability current through the system, $J_{\infty}$, is constant:
\begin{equation} 
\label{Jinfdef}
  \left[- D\frac{\partial }{\partial x} + \frac{F(x)}{\gamma}\right] p_{\rm st}(x) = J_{\infty}. 
\end{equation}
The stationary distribution $p_{\rm st }(x)$ is determined by the integration of this equation. 
\begin{equation}
\label{PDFstac}
p_{\rm st}(x) = - \frac{J_{\infty}}{D} 
{{\rm e}^{-\beta U(x) } \int_{-\infty}^{x} dy \, {\rm e}^{\beta U(y) }}. 
\end{equation} 
It follows from the normalization condition that the stationary current is related to the MFPT~(\ref{MFPTinfinf}) as follows,
\begin{equation}
\label{Jinf} 
J_{\infty} = - \frac{1}{\left< {\bf T} (\infty \to -\infty)  \right>},
\end{equation}
where the minus sign tells us that the probability flows from right to left. The stationary distribution in the cubic potential is illustrated in Fig.~\ref{fig:PDFstac} for different temperatures. Notice that the tail of the stationary PDF when $|x| \to \infty$ reads
\begin{equation} 
p_{\rm st}(x) \approx - \frac{J_{\infty}}{D}
\frac{k_B T}{n \mu |x|^{n-1}}. 
\end{equation} 
Thus the PDF in the steady state has no moments when the potential is cubic.  They can be obtained from simulations or experiments only for the fixed interval $x\in (a,y)$ between the boundaries. In principle, they may be used to visualize thermally-induced effects, but they cannot be compared with the steady-state distribution (\ref{PDFstac}). Instead of computing the moments, it is more convenient to analyze the position of the maximum of the distribution $p_{st}(x)$. As $T\to 0$ (low-noise regime), the stationary PDF is peaked near the origin $x=0$ (when there is no thermal noise, the particle approaches $x=0$ and remains there for a long time.). As the temperature $T$ increases, the peak shifts to the right towards higher values of the potential. At the same time, the width of the PDF increases. The position of the maximum is given by
\begin{equation}
\label{positionofmaximum}
x_{max} = C \left( \frac{k_{B}T}{\mu}\right)^{1/n}, 
\end{equation} 
where the constant $C$ is determined from the condition $dp_{\rm st}/dx=0$, 
\begin{equation}
C^{1-n}=  \int_{-\infty}^{C^{n}} d\xi \, {\rm e}^{\xi}\,\xi^{1/n-1} . 
\end{equation}
This counter-intuitive shift of the maximum of the stationary probability against the direction of the stationary particle current cannot be explained by a smaller fraction of MFPTs \ref{ratioMFPTs3} for $x>0$, because this ratio is temperature independent. When temperature is higher, two effects cooperate and result in this shift. Both are results of higher thermal energy of the surroundings. First, the particle can climb to higher value of the potential. Second, the transition through inflection point is much faster for larger $T$, hence trajectory spends less time in the vicinity of the inflection point. 

In experiment, the position of the maximum is a quantity which can be determined from a small sample of trajectories. Moreover, when observing the maximum we do not need to sample the complete PDF, whose tails are heavy and are determined by fast diverging trajectories.

%%%%%%%%%%%%%%%%%%%%%%%%%%%%%%%%%%%%%%%%%%%%%%%%%%%%%%%%%%%%%%%%%%%%%%%%%%%%%%%%%%%%%%%%%%%%%%%%%%%%%%%%%%%
%%%%%%%%%%%%%%%%%%%%%%%%%%%%%%%%%%%%%%%%%%%%%%%%%%%%%%%%%%%%%%%%%%%%%%%%%%%%%%%%%%%%%%%%%%%%%%%%%%%%%%%%%%%
\section{Conclusion}
%%%%%%%%%%%%%%%%%%%%%%%%%%%%%%%%%%%%%%%%%%%%%%%%%%%%%%%%%%%%%%%%%%%%%%%%%%%%%%%%%%%%%%%%%%%%%%%%%%%%%%%%%%%
%%%%%%%%%%%%%%%%%%%%%%%%%%%%%%%%%%%%%%%%%%%%%%%%%%%%%%%%%%%%%%%%%%%%%%%%%%%%%%%%%%%%%%%%%%%%%%%%%%%%%%%%%%%

We have analyzed thermally-induced stochastic overdamped dynamics of a Brownian particle in the unstable potentials $U(x)=\mu x^n$. The main aim of this paper was to provide exact results for the quantities, which can be easily accessible in the experiment. The problem with the unstable potential $U(x)=\mu x^n$ is that the stochastic trajectories are able to escape to minus infinity at a finite time. This severely restricts validity of the transient analysis based on the moments of the particle position \cite{RadimPavelI}. Instead, we propose to characterize the particle transient dynamics using the first-passage properties. Such approach eliminates diverging trajectories and yields results (MFPT, SNR, and universal fractions of MFPTs) which are easily accessible and could be reliably determined even from a small  sample of trajectories (in numerical simulations we have used $N_{traj}=1000$).  

In contrast to the linear potential $U(x)=\mu x$, where the MFPT does not depend on the temperature, we observe shortening of the MFPT in the nonlinear case. We have derived the upper bound for the MFPT for a given temperature, Eq.~(\ref{MFPTinfinf}). This formula, valid when $y$ and $-a$ are large as compared to thermal length scale $l$, dictates the temperature-dependence of MFPT. It should be corrected by substracting the temperature-independent term, see Eq.~(\ref{MFPTxnytoa}), when $y$, $a$ are large but finite. We have also shown that ratios of MFPTs, Eqs.~(\ref{ratioMFPTs3}),~(\ref{ratioMFPTs5}) are very good and universal identificators of the underlying nonlinear processes. They depend solely on the power of the nonlinear potential $n$ and not on the temperature, strength of nonlinearity and damping.

In the second part of the paper, we have shown how to characterize position of the particle in the unstable optical potential. To this end we propose a cyclic experiment with a help of fast return mechanics. In this case, easily measurable quantity, which reflects all essential features of the nonlinear potential, is the position of the maximum $x_{max}$ of the non-equilibrium stationary distribution. For higher temperatures, this maximum climbs towards higher values of the potential in the opposite direction as compared to the particle current. Again, in order to measure this quantity, we do not need the precise information about very fast diverging trajectories since the maximum of the PDF lies in the region of slow noisy dynamics. 

All these thermally-induced effects are an interesting targets for the experimental investigation in the current version of optical tweezers. They will allow much deeper investigation of more complex thermally-induced effects in the nonlinear stochastic dynamics.          

An interesting direction for future work is to study an extension of the cubic nonlinearity and other highly nonlinear mechanical systems to  quantum regime, where thermodynamical aspects are subject of current research \cite{Berut2012, M1, L1, X1}. Another immensely interesting problem is to describe an influence of a nonlinear potential on anomalous diffusion \cite{Klafter2005}, such as that occurring in crowded biological environments \cite{Franosch2013}.\\

%%%%%%%%%%%%%%%%%%%%%%%%%%%%%%%%%%%%%%%%%%%%%%%%%%%%%%%%%%%%%%%%%%%%%%%%%%%%%%%%%%%%%%%%%%%%%%%%%%%%%%%%%%%
%%%%%%%%%%%%%%%%%%%%%%%%%%%%%%%%%%%%%%%%%%%%%%%%%%%%%%%%%%%%%%%%%%%%%%%%%%%%%%%%%%%%%%%%%%%%%%%%%%%%%%%%%%%
\section*{Acknowledgments}
%%%%%%%%%%%%%%%%%%%%%%%%%%%%%%%%%%%%%%%%%%%%%%%%%%%%%%%%%%%%%%%%%%%%%%%%%%%%%%%%%%%%%%%%%%%%%%%%%%%%%%%%%%%
%%%%%%%%%%%%%%%%%%%%%%%%%%%%%%%%%%%%%%%%%%%%%%%%%%%%%%%%%%%%%%%%%%%%%%%%%%%%%%%%%%%%%%%%%%%%%%%%%%%%%%%%%%%
R.F.\ and P.Z.\ acknowledge support from the Czech Science Foundation (project GB14-36681G).

%%%%%%%%%%%%%%%%%%%%%%%%%%%%%%%%%%%%%%%%%%%%%%%%%%%%%%%%%%%%%%%%%%%%%%%%%%%%%%%%%%%%%%%%%%%%%%%%%%%%%%%%%%%
%%%%%%%%%%%%%%%%%%%%%%%%%%%%%%%%%%%%%%%%%%%%%%%%%%%%%%%%%%%%%%%%%%%%%%%%%%%%%%%%%%%%%%%%%%%%%%%%%%%%%%%%%%%
\bibliography{notesFPT}

%%%%%%%%%%%%%%%%%%%%%%%%%%%%%%%%%%%%%%%%%%%%%%%%%%%%%%%%%%%%%%%%%%%%%%%%%%%%%%%%%%%%%%%%%%%%%%%%%%%%%%%%%%%
%%%%%%%%%%%%%%%%%%%%%%%%%%%%%%%%%%%%%%%%%%%%%%%%%%%%%%%%%%%%%%%%%%%%%%%%%%%%%%%%%%%%%%%%%%%%%%%%%%%%%%%%%%%
\end{document}